\documentclass[iop]{emulateapj}

\def\Msun{M_\odot}
\def\Mjup{M_J}

\begin{document}

\title{The PHASES Differential Astrometry Data Archive.  V.  Candidate Substellar Companions to Binary 
Systems}

\author{Matthew W.~Muterspaugh\altaffilmark{1, 2}, 
Benjamin F.~Lane\altaffilmark{3}, S.~R.~Kulkarni\altaffilmark{4}, 
Maciej Konacki\altaffilmark{5, 6}, Bernard F.~Burke\altaffilmark{7}, 
M.~M.~Colavita\altaffilmark{8}, M.~Shao\altaffilmark{8}, 
William I. Hartkopf\altaffilmark{9}, Alan P.~Boss\altaffilmark{10}, 
M.~Williamson\altaffilmark{2}}
\altaffiltext{1}{Department of Mathematics and Physics, College of Arts and 
Sciences, Tennessee State University, Boswell Science Hall, Nashville, TN 
37209 }
\altaffiltext{2}{Tennessee State University, Center of Excellence in 
Information Systems, 3500 John A. Merritt Blvd., Box No.~9501, Nashville, TN 
37209-1561}
\altaffiltext{3}{Draper Laboratory,  555 Technology Square, Cambridge, MA 
02139-3563}
\altaffiltext{4}{Division of Physics, Mathematics and Astronomy, 105-24, 
California Institute of Technology, Pasadena, CA 91125}
\altaffiltext{5}{Nicolaus Copernicus Astronomical Center, Polish Academy of 
Sciences, Rabianska 8, 87-100 Torun, Poland}
\altaffiltext{6}{Astronomical Observatory, Adam Mickiewicz University, 
ul.~Sloneczna 36, 60-286 Poznan, Poland}
\altaffiltext{7}{MIT Kavli Institute for Astrophysics and Space Research, 
MIT Department of Physics, 70 Vassar Street, Cambridge, MA 02139}
\altaffiltext{8}{Jet Propulsion Laboratory, California Institute of 
Technology, 4800 Oak Grove Dr., Pasadena, CA 91109}
\altaffiltext{9}{U.S.~Naval Observatory, 3450 Massachusetts Avenue, NW, Washington, DC, 20392-5420}
\altaffiltext{10}{Department of Terrestrial Magnetism, Carnegie Institution of
Washington, 5241 Broad Branch Road, NW, Washington, DC 20015-1305}

\email{matthew1@coe.tsuniv.edu, blane@draper.com, maciej@ncac.torun.pl}

\begin{abstract}
The Palomar High-precision Astrometric Search for Exoplanet Systems 
monitored 51 subarcsecond binary systems to evaluate whether tertiary 
companions as small as Jovian planets orbited either the primary or secondary 
stars, perturbing their otherwise smooth Keplerian motions.  Six binaries are 
presented that show evidence of substellar companions orbiting either the 
primary or secondary star.  Of these six systems, the likelihoods of two of 
the detected perturbations to represent real objects 
are considered to be ``high confidence'', while the remaining four systems are 
less certain and will require continued observations for confirmation.  
\end{abstract}


\section{Introduction}

The use of astrometric measurements to detect the reflex motions of stars 
caused by substellar companions orbiting them has a long 
history filled with false alarms.  Famously, \cite{vanDeKamp1963} 
claimed to have discovered giant planet companions to Barnard's Star.  
His first estimates of a single planet of 1.6 times the mass of 
Jupiter with an orbital period of 24 years and an eccentricity of 0.6 
were later revised 
to two planets with orbital masses and periods of 1.1 $\Mjup$ at 26 years and 
0.8 $\Mjup$ at 12 years \citep{vanDeKamp1969}.  He never accepted growing 
evidence from other astronomers that these 
discoveries were not repeatable elsewhere \citep{Gatewood1973, hershey1973}; 
today it has been shown conclusively that these planets are not real 
\citep{2003A&A...403.1077K}.

\cite{Han2001} used {\em Hipparcos} 
measurements to analyze stars with known radial velocity (RV) detected 
exoplanet candidates.  The precision of {\em Hipparcos} was 
insufficient to detect the reflex motions if the objects are in fact 
planets; however, if instead 
the orbits are face-on, the actual companion masses would be larger than the 
RV-derived masses, so the resulting much larger motions could have been 
detected by {\em Hipparcos}.  This provided 
a test of whether the RV candidates were 
in fact mostly face-on binaries (transiting planets prove that at least some 
RV candidates are real planets---see, for example, \cite{Henry209458}---but 
this is not applicable to the vast majority of systems).  \cite{Han2001} 
concluded 
that most RV candidates {\em did} show orbital motions in the {\em Hipparcos} 
database, and were thus binary stars, not planetary systems.  However, 
\cite{Pourbaix2001} showed the 
orbital analysis to be incorrect, and a proper statistical 
analysis reveals no credible detections; instead the 
results are consistent with randomly 
oriented orbits and most of the RV-detected objects being planetary in 
nature.

A few RV-detected planetary systems have had their orbital geometries 
constrained by {\em Hubble Space Telescope} astrometry 
\citep{ben02b, McArthur2010}.  Though impressive work, 
these do not represent discoveries of new systems by astrometry, 
and the ratio of measurement precision to signal 
amplitude is low enough to make it unlikely the astrometry could have produced 
a detection by itself (or even have been made, given the time requirements of 
a blind search with {\em Hubble}) had the RV detection not already been 
present.  \cite{2005ApJ...630..528P} successfully used astrometry to discover 
a brown dwarf companion to a M dwarf, a promising first step on the 
path to finding true planets.  The brown dwarf was later confirmed by direct 
imaging \citep{2006ApJ...650L.131L}.  
Most recently, \cite{PravdoShaklan2009} claimed an astrometric 
detection of a giant planet around a nearby M dwarf from the STEPS project, 
using standard CCD astrometry from large aperture telescopes.  However, 
this candidate was rapidly shown to be inconsistent with RV observations by 
\cite{Bean2010}.

It is thus with some trepidation that we announce the candidate substellar 
companions orbiting either the primary or secondary stars in several binaries 
studied using differential astrometry by PHASES---the Palomar High-precision 
Astrometric Search for Exoplanet Systems.  Given that other 
astrometrically ``discovered'' substellar objects have not withstood the test 
of continued observations, these may represent either the first such companions 
detected, or the latest in the tragic history of this challenging approach.  

Given the challenges of astrometry, why would it be considered a preferred 
way to detect planets in current and future searches?  Astrometry has a number 
of advantages over other techniques:
\newcounter{countA}
\begin{list}{\arabic{countA}.}{\usecounter{countA}\setlength{\leftmargin}{0.2in}}
\item Astrometry and RV provide information about 
the masses of companions to nearby stars.  
Other methods are insensitive to this fundamental property.
  \begin{list}{\labelitemi  }{\setlength{\leftmargin}{0.1in}}
  \item Because the reflex motion of the star is monitored, the mass of the 
  companion is measured directly.  For nearby systems that can be followed up 
  by direct imaging, only RV and astrometry provide such information.
  \item The two-dimensional nature of the astrometric measurement provides 
  unique mass 
  estimates.  In contrast, RV detections only give the companion mass times 
  the sine of the unknown inclination, $M\sin i$.
  \end{list}
\item Astrometry is effective in regimes where RV has reduced precision:
  \begin{list}{\labelitemi  }{\setlength{\leftmargin}{0.1in}}
  \item Astrometry can operate over a wide range of stellar masses, rotational 
        velocities, and spectral types, to better explore relationships 
        between 
        the properties of the host star and its planetary system.  RV is 
        most effective for slowly rotating, mid-to-late type stars.
  \item Astrometric sensitivity increases with companion period, an opposite 
        trend as RV.  This is particularly important when 
        identifying long period planets for direct imaging work, where the 
        wider star-planet separation reduces technical challenges for imaging.
  \item Astrometry is less sensitive to surface vibrations and star spots than 
        RV \citep{makarov2009}.  
        This is particularly important for identifying the small 
        (1 $\mu {\rm as}$ and 0.1 ${\rm m\, s^{-1}}$) signals of Earthlike 
        planets in the habitable zones of nearby, Sunlike stars.  This 
        motivates future astrometric planet searches.
  \item Astrometry is well suited to studying planets in binary systems.  This 
        is the primary motivation for using the astrometric method for 
        PHASES.  RV can study planetary companions to a few binary systems.  
        For example, binaries with very large sky separations can be studied, 
        which frequently implies large physical separations as well, and 
        these evolve rather like single stars, revealing little new about 
        planetary system formation and evolution.

        It is instead binaries 
        with separations in the critical
        $\sim 10-50$ AU range that can greatly contribute new information.  
        This range is wide enough that planets can have stable orbits around
        either star if present, but close enough that the second star may 
        influence formation of the planet in the first place.  RV can study a 
        few special cases of these binaries:  those very close to the solar 
        system \citep[e.g., $\gamma$ Cep; ][]{Campbell1988, Hatzes2003}
        so the components are spatially resolved, a few  
        high contrast systems 
        \citep[such that the second star minimally impacts the spectrum; e.g., HD 126614;][]{howard2010}, 
        and a few triple star systems, where a short-period stellar 
        subsystem causes the spectral features to be split 
        \citep[e.g., the controversial companion to HD 188753;][]{Konacki2005}.  
        In other cases, the stars are often both spatially unresolved and 
        spectrally blended, making precision 
        velocities impossible; even when the lines can be separated, precision 
        RV on the double spectrum is challenging \citep{Konacki04}.  
  \end{list}
\end{list}

Thus, PHASES used astrometry to observe 51 binary systems with the goal of 
identifying new tertiary companions orbiting either of the bright stars in 
the system.  Of these, 33 systems have more than 10 successful observations, 
allowing a realistic chance for a companion search to be successful.  Seven of 
those 33 systems are either triple or quadruple stars, five more are so 
distant that their physical separations fall well outside of the 50 AU limit 
(two of these---HD 171779 and HD 221673---may have brown dwarf companions, and 
are presented in this paper), and six more have semimajor axes less than 10 
AU, though these last two classes are useful to verify the astrometric 
technique.  The remaining 15 systems provide a sample from which the frequency 
of planets in closely separated binaries can be 
evaluated.  \cite{PfahlMute2006} showed that stellar 
encounters, even in star forming regions where the stellar density is higher 
than typical space, are rare enough that only $\sim 0.1\%$ of closely 
separated binaries could pick up a planet that had not originally formed as a 
companion in the binary itself but rather via an exchange or binary hardening 
event.  Observed frequencies of planets in close 
binaries that are higher than this value offer 
evidence of in situ formation.  If giant 
planets do form in these binaries, it is likely the process must be rapid.  
Current core-accretion models predict slow formation, though the competing 
gravitational instability method shows promise at rapid formation.  Thus, the 
frequency of planet formation in these close binaries evaluates the relative 
frequencies with which these (and other) modes of giant planet formation occur 
in nature.

Unfortunately, the statistics of the number of binaries that have been 
observed by RV are difficult to evaluate.  However, several planets have been 
found in close binaries by RV methods, certainly more than the 
$0.1\%$ frequency predicted by non-in situ formation; see Table 
\ref{tab:tabclose}.  
The next challenge is to evaluate the planet frequency in a less biased 
manner.  Though limited in size, the PHASES sample represents an attempt to 
contribute to this effort.

\begin{deluxetable*}{lcccccc}
\tablecolumns{7}
\tablecaption{Close Binaries with Substellar Companions. \label{tab:tabclose}}
\tablehead{
\colhead{System} &
\colhead{Object Type\tablenotemark{a}} &
\colhead{$a$(AU)} &
\colhead{$e$\tablenotemark{b}} &
\colhead{$M_1/M_2$\tablenotemark{c}} &
\colhead{$R_t$(AU)\tablenotemark{d}} &
\colhead{References}
}
\startdata
$\gamma$ Cephei             & p  & 18.5     & 0.36    & 1.59/0.34   & 3.6       & 1, 2 \\
GJ 86 \tablenotemark{e}     & p  & $\sim$20 & \nodata & 0.7/1.0     & $\sim$5   & 3, 4 ,5 \\
HD 41004                    & p  & $\sim$20 & \nodata & 0.7/0.4     & $\sim$6   & 6 \\
HD 41004                    & bd & $\sim$20 & \nodata & 0.4/0.7     & $\sim$5   & 6 \\
HD 126614                   & p  & $\sim$45 & \nodata & 1.145/0.324 & $\sim$15  & 7 \\
HD 188753\tablenotemark{f}  & p  & 12.3     & 0.50    & 1.06/1.63   & 1.3       & 8, 9, 10 \\
HD 196885                   & p  & $\sim$25 & \nodata & 1.3/0.6     & $\sim$8   & 11 \\
HD 176051                   & p  & 19.1     & 0.2667  & 0.71/1.07   & 3.2       & This work \\
HD 221673                   & bd & 95       & 0.322   &   2/2       & 12.6      & This work 
\enddata
\tablenotetext{a}{``p'' indicates a giant planet companion and 
``bd'' indicates the companion is a brown dwarf.}
\tablenotetext{b}{When the eccentricity is unknown, 
the projected binary separation is used as an approximation, except 
in the case of HD 126614, where a linear velocity 
trend is observed, and the binary itself has been 
resolved, leading to two possible solutions with 
$a = 40^{+7}_{-4}$ and $50^{+2}_{-3}$ AU.}
\tablenotetext{c}{Mass of star hosting planet divided 
by mass of the companion star (in solar masses).}
\tablenotetext{d}{The distance from the primary 
star at which a disk would be rapidly 
truncated by tides \citep{Pichardo2005}.}
\tablenotetext{e}{The companion star is a 
white dwarf of mass $\simeq$$0.5\Msun$.  
To estimate $R_t$ at the time of formation, 
an original companion mass of $1\Msun$ is assumed.}
\tablenotetext{f}{The companion star itself is a 
binary with semimajor axis 0.67\,AU.  
This candidate is controversial due to minimal 
data in the discovery paper with sporadic 
observing cadence and a lack of evidence found
 by \cite{2007A&A...466.1179E} and \cite{2009MNRAS.399..906M}.}
\tablerefs{
(1) Campbell et al.~1988; \nocite{Campbell1988}
(2) Hatzes et al.~2003; \nocite{Hatzes2003}
(3) Queloz et al.~2000; \nocite{Queloz2000} 
(4) Mugrauer \& Neuh{\"a}user 2005; \nocite{Mugrauer2005} 
(5) Lagrange et al.~2006; \nocite{Lagrange2006}
(6) Zucker et al.~2004; \nocite{Zuc2004} 
(7) Howard et al.~2010; \nocite{howard2010}
(8) \cite{Konacki2005}; 
(9) Eggenberger et al.~2007; \nocite{2007A&A...466.1179E}
(10) Mazeh et al.~2009; \nocite{2009MNRAS.399..906M}
(11) Chauvin et al.~2006 \nocite{Chauvin2006}
}
\end{deluxetable*}

This paper is the fifth in a series, analyzing the final results of the PHASES 
project after its completion in late 2008.  The first paper describes the 
observing method, sources of measurement uncertainties, limits of observing 
precisions, derives empirical scaling rules to account for noise sources 
beyond those predicted by the standard reduction algorithms, and presents the 
full catalog of astrometric measurements from PHASES \citep{Mute2010A}.  
The second paper combines PHASES astrometry with astrometric measurements 
made by other methods as well as RV observations (when 
available) to determine orbital solutions to the binaries' Keplerian motions, 
determining physical properties such as component masses and system distance 
when possible \citep{Mute2010B}.  The third paper presents limits on the 
existence of substellar tertiary companions, orbiting either the primary or 
secondary stars in those systems, that are found to be consistent with being 
simple binaries \citep{Mute2010C}.  The fourth paper presents three-component 
orbital solutions to a known triple star system (63 Gem A $=$ HD 58728) and a 
newly discovered triple system (HR 2896 $=$ HD 60318) \cite{Mute2010D}.  
Finally, the current paper presents candidate substellar companions to PHASES 
binaries as detected by astrometry.

Astrometric measurements were made 
as part of the PHASES program at the Palomar Testbed 
Interferometer \citep[PTI;][]{col99}, which was located on Palomar Mountain 
near San Diego, California.  It was developed by the Jet 
Propulsion Laboratory, California Institute of Technology for NASA, as a 
testbed for interferometric techniques applicable to the Keck Interferometer 
and other missions such as the Space Interferometry Mission (SIM).  It 
operated in the J ($1.2 \mu{\rm m}$), H ($1.6 \mu{\rm m}$), and K 
($2.2 \mu{\rm m}$) bands, and combined starlight from two out of three 
available 40 cm apertures.  The apertures formed a triangle with one 110 m and 
two 87 m baselines.  PHASES observations began in 2002 continued through 
2008 November when PTI ceased routine operations.

\section{Algorithm for Identifying Astrometric Companions}

Blind searches were conducted to identify potential tertiary companions to 
the PHASES binaries.  
An algorithm based on that of \cite{cumming1999} and 
\cite{cumming2008} was modified for use on astrometric 
data for binary systems, 
as described in Paper III, and used to conduct blind searches for 
tertiary companions in these systems.

The overall procedure is to create a periodogram of an 
F statistic comparing the goodness-of-fit $\chi^2$ between a single Keplerian 
model and that for a double Keplerian model for a number of possible orbital 
periods for the second orbit.  The orbital periods selected were chosen to be 
more than Nyquist sampled, to ensure complete coverage, as $P = 2 f T / k$
where $T$ is the span of PHASES observations, $f=3$ is an oversampling factor, 
and $k$ is a positive integer.  Two searches were conducted for each 
binary:  first, with the use of only the PHASES measurements, and 
second with both the PHASES and non-PHASES astrometry, to better constrain 
the wide binary motion during the search.  In addition to the positive 
integer values of $k$, the period 
corresponding to $k=1/2$ was evaluated to search for companions with orbits 
slightly longer than the PHASES span.

The orbital period for which the F statistic periodogram has its maximum 
value is 
the most likely orbital period of a companion object.  To ensure the peak is 
a real object rather than a statistical fluctuation, 1000 synthetic data sets 
with identical cadence and measurement uncertainties as the actual data were 
created and evaluated in the same manner.  The fraction of these having a 
maximum F statistic larger than that of the actual data provided an estimate 
of the false alarm probability (FAP) that the signal is not caused by an 
actual companion.

\section{PHASES Measurements}

PHASES differential astrometric measurements were obtained with the observing 
method and standard data analysis pipeline described in Paper I.  The 
measurements themselves and associated measurement uncertainties are also 
tabulated in Paper I.  The number of 
PHASES measurements available for each of the six systems being 
investigated are listed in Table \ref{tab::speckleUnitUncert}.

\vspace{0.5in}
\section{Non-PHASES Astrometry}

Measurements of binaries observed by PHASES made by previous astrometric 
techniques and cataloged in the {\em Washington Double Star Catalog} 
\citep[WDS][]{wdsCatalog, wdsCatalogUpdate} 
were assigned weights according to the formula described 
by \cite{hart01}.  These allowed refined planet searches in which the binary 
orbit itself is better constrained by the longer duration, though lower 
precision, astrometric measurements.  Including these measurements lifts some 
degeneracies in the double-orbit modeling.

Unit weight uncertainties in separation and position 
angle were evaluated by the following iterative procedure.  First guess 
values for the unit uncertainties of 24 mas in separation and $1.8^\circ$ 
in position angle were assigned to the measurements of a given binary; these 
values corresponded to previous experience using this procedure on $\mu$ Ori 
\citep{MuteMuOri2008}.  Second, the measurements were fit to a Keplerian model 
and the orbital parameters were optimized to minimize the fit $\chi^2$.  
This intrinsically assumes the non-PHASES astrometric measurements are 
insensitive to the tertiary companions being sought, an assumption that will 
be justified given the small sizes of the perturbations detected.  
Third, the weighted scatter of the residuals in separation and position angle 
were evaluated, and the guessed unit uncertainties updated to make the rms 
scatter in each equal to unity.  Fourth, the second and third steps were 
iterated two more times, at which point the values converged.  Fifth, the 
final unit uncertainties were multiplied by the square root of the reduced 
$\chi^2$ ($\sqrt{\chi_r^2}$) of the fit, and refit one more time with these 
slightly larger weights.  Sixth, if no residuals deviated by more than 
3$\sigma$, the process ended, otherwise, the single measurement with the 
largest separation or 
position angle residual (weighted by its uncertainty) was flagged as an 
outlier, and removed from future fits.  Seventh, the process was repeated at 
the first step.  The resulting weights are listed in Table 
\ref{tab::speckleUnitUncert} and the measurements themselves are listed in 
Table \ref{tab::speckleData}.

\begin{deluxetable}{lllllll}
\tablecolumns{7}
\tablewidth{0pc} 
\tablecaption{Number of PHASES and Non-PHASES Measurements and Unit Weight Uncertainties for Non-PHASES Measurements \label{tab::speckleUnitUncert}}
\tablehead{ 
\colhead{HD Number} & \colhead{${\rm N_{P}}$} & \colhead{${\rm N_{P, O}}$} & \colhead{${\rm N_{NP}}$} & \colhead{${\rm N_{NP, O}}$} & \colhead{$\sigma_{\rho, \circ}$} & \colhead{$\sigma_{\theta, \circ}$}}
\startdata
13872  & 89 &  0 & 103 & 14 & 0.013 & 2.51 \\
171779 & 54 &  0 & 128 & 12 & 0.020 & 2.79 \\
176051 & 65 &  1 & 327 & 12 & 0.140 & 4.48 \\
196524 & 72 &  1 & 598 & 48 & 0.046 & 2.78 \\
202444 & 39 &  0 & 286 & 13 & 0.123 & 4.37 \\
221673 & 98 &  1 & 333 & 21 & 0.056 & 2.12 
\enddata
\tablecomments{
The numbers of PHASES and non-PHASES astrometric measurements used for orbit 
fitting with each of the binaries being studied are presented in Columns 2 and 
4 respectively, along with the additional 
numbers of measurements rejected as outliers 
in Columns 3 and 5.  Columns 6 and 7 list the 1$\sigma$ measurement 
uncertainties for unit weight measurements from 
non-PHASES observations determined by iterating Keplerian fits to the 
measurements with removal of 3$\sigma$ or greater outliers in either 
dimension.  Columns 6 and 7 are in units of arcseconds and degrees, 
respectively.
}
\end{deluxetable}

\begin{deluxetable*}{llllllll}
\tablecolumns{8}
\tablewidth{0pc} 
\tablecaption{Non-PHASES Astrometric Measurements \label{tab::speckleData}}
\tablehead{ 
\colhead{HD Number} & \colhead{Date} & \colhead{$\rho$} & \colhead{$\theta$} 
& \colhead{$\sigma_{\rho}$} & \colhead{$\sigma_{\theta}$} & \colhead{Weight} 
& \colhead{Outlier} \\
\colhead{} & \colhead{(year)} & \colhead{(arcsec)} & \colhead{(deg)} 
& \colhead{(arcsec)} & \colhead{(deg)} & \colhead{} 
& \colhead{}}
\startdata
13872 & 1965.9100 & 0.250 & 49.70 & 0.015 & 2.81 & 0.8 & 1 \\
13872 & 1966.1100 & 0.230 & 50.70 & 0.029 & 5.61 & 0.2 & 0 \\
13872 & 1966.7200 & 0.220 & 42.80 & 0.012 & 2.29 & 1.2 & 0 \\
13872 & 1967.0699 & 0.190 & 39.50 & 0.016 & 3.00 & 0.7 & 0 \\
221673 & 2006.7170 & 0.560 & 98.20 & 0.102 & 3.87 & 0.3 & 0 \\
221673 & 2006.9750 & 0.554 & 97.70 & 0.030 & 1.13 & 3.5 & 0 \\
221673 & 2007.2800 & 0.580 & 99.20 & 0.040 & 1.50 & 2.0 & 0 \\
221673 & 2008.8850 & 0.540 & 100.45 & 0.018 & 0.69 & 9.5 & 0 
\enddata
\tablecomments{
Non-PHASES astrometric measurements from the WDS Catalog 
are listed with 1$\sigma$ measurements uncertainties, and weights.
Column 1 is the HD catalog number of the target star, Column 2 is 
the decimal year of the observation, Columns 3 and 4 are the separation in 
arcseconds and position angle in degrees, respectively, Columns 5 and 6 are 
the 1-$\sigma$ uncertainties in the measured quantities from Columns 3 and 4, 
Column 7 is the weight assigned to the measurement, and Column 8 is 1 if the
measurement is a $>$3$\sigma$ outlier and omitted from the fit, 0 otherwise.  
(This table is available in its entirety in machine-readable 
and Virtual Observatory (VO) forms in the online journal.  A portion is shown 
here for guidance regarding its form and content.)
}
\end{deluxetable*}

\vspace{1.0in}
\section{Substellar Companions with High Degrees of Confidence}

\subsection{HD 176051}

HD 176051 (HR 7162, HIP 93017, WDS 18570$+$3254, and, though rarely used, the 
proper name of Inrakluk has been proposed) is an intriguing PHASES 
binary because its components are relatively low mass (1.07 and 0.71 $\Msun$), 
and the system is relatively nearby ($14.99 \pm 0.13$ pc) as determined by 
{\em Hipparcos} observations \citep[][hereafter S99]{Soder1999}.  
Both of these qualities 
indicate astrometric perturbations by 
tertiary companions will have relatively large signals.  The model of 
\cite{holman1999} (hereafter HW99) for determining which planetary 
orbits in binary systems are stable long-term 
predicts companions with periods as long as $\sim 3000$ days 
can have stable orbits.  

\begin{figure*}[!ht]
\plottwo{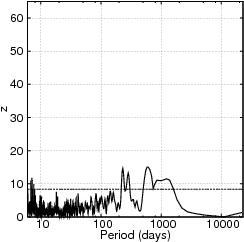}{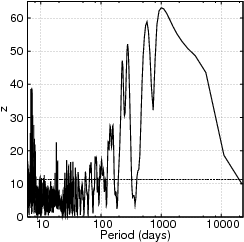}
\caption[Periodograms for Tertiary Companions to HD 176051 (HR 7162)]
{ \label{fig::periodograms_176051}
Periodograms of the F statistic comparing models with and without tertiary 
companions to HD 176051 (HR 7162) in astrometric-only models.  The left 
figure is for analysis only using the PHASES data, while the right is for 
combined analysis of PHASES and non-PHASES astrometric measurements.
For HD 176051 the 1\% FAP is at $z=8.37$ for the PHASES-only analysis, and 
$z=11.33$ for the combined analysis, as indicated by horizontal lines.
}
\end{figure*}

The initial PHASES-only search for companions found a most significant peak 
of $z = 15.0$ at a period of 581 days with FAP 0.0\%.  
This low value inspired a 
revised search including both the PHASES and lower-precision non-PHASES 
astrometric observations 
to investigate whether the detection continued to be valid.  The revised 
search finds the most significant peak of $z=63.1$ at a period of 1004 days 
with FAP 0.0\%.  These two distinct orbital periods are 
probably aliases of each other and 
both are present in both periodograms; see Figure 
\ref{fig::periodograms_176051}.

The two peaks may indicate aliasing, 
orbital eccentricity, or confusion with the wide binary orbit.  Both 
companion orbital periods were further explored using a double Keplerian model 
optimizing all orbital elements, including the companion orbital period and 
allowing for non-circular companion orbits.  Both the 
fit $\chi^2$ and visual inspection of the orbital 
solution confirmed that the longer period solution is more likely to be 
correct and the other is a harmonic.  Additionally, 
smaller peaks corresponding to 276, 
225, and 7 days were also explored, but did not produce convincing solutions 
at all.  

The eccentricity of the subsystem orbit is not constrained by the astrometry 
measurements.  This is likely due to the PHASES measurements being high 
precision only in one dimension, making it difficult to use Kepler's second 
law to lift ambiguity between inclined circular orbits and face-on eccentric 
ones.  The eccentricity is fixed at zero in the present analysis.  

\begin{deluxetable}{lll}
\tablecolumns{3}
\tablewidth{0pc} 
\tablecaption{Orbit Model for HD 176051\label{tab::hd176051_parameters}}
\tablehead{
\colhead{Parameter} & \colhead{Value} & \colhead{Uncertainty} 
}
\startdata 
$P_{A-B}$ (days)          & 22430   & 15      \\
$T_{A-B}$ (MHJD)          & 41384   & 23      \\
$e_{A-B}$                 & 0.2667  & 0.0022  \\
$a_{A-B}$ (arcsec)    & 1.2756  & 0.0023  \\ 
$i_{A-B}$ (deg)       & 114.159 & 0.078   \\
$\omega_{A-B}$ (deg)  & 281.71  & 0.26    \\
$\Omega_{A-B}$ (deg)  & 48.846  & 0.093   \\
$P_{\rm Ba-Bb}$ (days)     & 1016    & 40      \\
$T_{\rm Ba-Bb}$ (MHJD)     & 53583   & 39       \\
$e_{\rm Ba-Bb}$            & 0       & (Fixed)  \\
$a_{COL}$ (${\rm \mu as}$) & 241  & 41  \\ 
$i_{\rm Ba-Bb}$ (deg)  & 115.8   & 8.2      \\                        
$\omega_{\rm Ba-Bb}$ (deg) & 0  & (Fixed)  \\
$\Omega_{\rm Ba-Bb, 1}$ (degrees) &  69 & 11   \\
$\chi^2$ and dof          & 1015.3  & 772     
\enddata
\tablecomments{
Best fit orbital elements in the Campbell basis for HD 176051, with $1\sigma$ 
uncertainties.
}
\end{deluxetable}

\begin{figure}[!ht]
\plotone{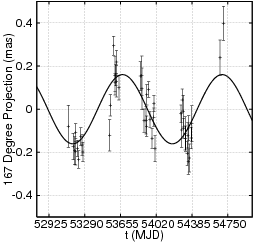}
\caption[Center-Of-Light Orbit of the HD 176051 Subsystem]
{ \label{fig::hd176051_reflex}
Time series of PHASES observations of HD 176051 (HR 7162) for the 1016 day 
subsystem orbital solution, 
measured along an axis at angle $167^\circ$ from 
increasing differential right ascension through increasing differential 
declination; it was along this axis that the PHASES measurements are typically 
most sensitive.  For clarity, only measurements with uncertainties along this 
axis of $100\, {\rm \mu as}$ along this axis are shown.  
}
\end{figure}

The best 
fit Keplerian stellar binary$+$circular subsystem 
orbital solution is presented in Table 
\ref{tab::hd176051_parameters} and Figure \ref{fig::hd176051_reflex}.
The substellar object is a planet $1.5 \pm 0.3$ times the mass of 
Jupiter, assuming a distance of 15 pc and a stellar mass 
of $0.71 \, \Msun$, both based on the {\em Hipparcos} analysis by 
S99.  If the planet is instead around the more massive star, the 
planet's mass would be twice as large.

Interestingly, the binary and planetary orbits may be nearly 
coplanar.  The mutual inclination $\Phi$ of two orbits is given by
\begin{equation}\label{V819HerMutualInclination}
\cos \Phi = \cos i_1 \cos i_2  + \sin i_1 \sin i_2 \cos\left(\Omega_1 - \Omega_2\right)
\end{equation}
\noindent where $i_1$ and $i_2$ are the orbital inclinations and $\Omega_1$ 
and $\Omega_2$ are the longitudes of the ascending nodes.  When RV 
measurements are not available, there exists ambiguity in which node is 
ascending, and two different values of the mutual inclination are possible 
(corresponding to $\Omega_1 - \Omega_2$ varying by $180^\circ$).  In this case, 
the two possibilities are $\Phi = 18 \pm 17$ degrees or $\Phi = 126.4 \pm 6.6$ 
degrees.

\subsection{HD 221673}

HD 221673 (72 Peg, HR 8943, HIP 116310, WDS 23340$+$3120) is a pair of mid K 
giants.  \cite{baize1962} flagged it as possibly containing a variable star 
based on the scatter in the differential magnitude measurements by various 
observers.  However, {\em Hipparcos} photometry shows a scatter of only 6 
mmag.  The revised {\em Hipparcos} based parallax is $5.94 \pm 0.45$ mas 
\citep{vanleeuwen2007}.  This parallax and the best 
fit single Keplerian model predict an average stellar mass of 2 $\Msun$.  
Its long orbital period ($\sim 800$ years) implies a large range of orbits in 
which companions can be stable---up to 47 years according to the criteria of 
HW99.  Thus, binary dynamics are expected to have a smaller impact 
on planet formation in this system than others.

HD 221673 is extremely bright at infrared 
wavelengths ($K = 1.76$), is observable for long stretches 
during the late summer/fall months of best weather at Palomar, and served as 
one of the easiest and most reliable PHASES targets to observe.  As a result, 
98 PHASES measurements were successfully taken of HD 221673.  Despite there 
being large amounts of data available, single Keplerian orbit fitting was 
frustrated from the early beginnings---the data show much more scatter than 
predicted by the measurement uncertainties.

\begin{figure*}[!ht]
\plottwo{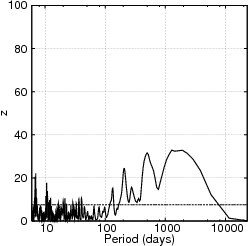}{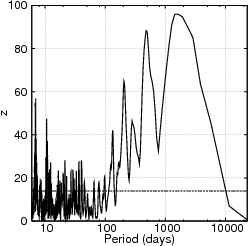}
\caption[Periodograms for Tertiary Companions to HD 221673 (72 Peg)]
{ \label{fig::periodograms_221673}
Periodograms of the F statistic comparing models with and without tertiary 
companions to HD 221673 (72 Peg) in astrometric-only models.  The left 
figure is for analysis only using the PHASES data, while the right is for 
combined analysis of PHASES and non-PHASES astrometric measurements.
For HD 221673 the 1\% FAP is at $z=7.56$ for the PHASES-only analysis, and 
$z=13.96$ for the combined analysis, as indicated by horizontal lines.
}
\end{figure*}

The initial PHASES-only search for companions found a most significant peak 
of $z = 32.9$ at a period of 1276 days ($k = 9$) with FAP 0.0\%.  
This low value inspired a 
revised search including both the PHASES and lower-precision non-PHASES 
astrometric observations 
to investigate whether the detection continued to be valid.  The revised 
search finds a most significant peak of $z=95.9$ at a period of 1435 days 
($k = 8$, within one sample of the peak value for the PHASES-only search) 
with FAP 0.0\%.  The periodograms are plotted in Figure 
\ref{fig::periodograms_221673}.

A few single-component RV measurements of 72 Peg have been published by 
\cite{tok2002} and \cite{abt1980}.  However, the relatively 
small number and short time coverage of each RV data set 
reduces any impact they have on 
constraining either the binary orbit or confirming the existence of additional 
components (especially low mass, long period companions).  Neither set shows 
variation in the velocities and as a result is not used in the 
orbital analysis.

A refined fit to the astrometric measurements was made for each of the three 
most significant peaks in the periodogram---the companion orbital period 
was seeded with values of 1435, 478, and 205 days.  Both circular and full 
Keplerian models were attempted as the companion orbit for each of 
the three periods being explored.  The longest period corresponded to the 
best fit of the three, though the companion's eccentricity was not 
constrained.  The best fit circular model converged with an orbital period of 
1539 days and $\chi^2 = 1549.2$, with 850 degrees of freedom, and is 
presented in Table \ref{tab::hd221673_parameters} and 
Figure \ref{fig::periodograms_221673}.  While $\chi^2$ 
is larger than the number of degrees of freedom, it is significantly improved 
compared to the model without a tertiary companion, for which $\chi^2 = 2669.9$ 
with 855 degrees of freedom.  The 1539 day companion is a brown dwarf with 
$35 \pm 4$ times the mass of Jupiter.

However, the remaining scatter and presence of other peaks in the periodogram 
(especially that at 478 days) leads one to question whether adding yet another 
Keplerian representing a fourth component to the system would yet further 
improve the fit.  A 3-Keplerian fit was seeded with the best parameters from 
the three-component, 2-Keplerian model, as well as the best orbit for a 478 
day companion as determined by the initial search.  First, both subsystem 
orbits 
were assumed to be circular.  This led to an improvement in the fit from 
$\chi^2 = 1549.2$ with 850 degrees of freedom for the 2-Keplerian model to 
$\chi^2 = 1422.9$ with 845 degrees of freedom for that with three 
Keplerians summed 
by superposition.  This is only a modest improvement, and at this time no 
detection is claimed for a second unseen object.  Alternatively, the 
remaining scatter could be 
related to the 6 mmag photometric variability.

\begin{deluxetable}{lll}
\tablecolumns{3}
\tablewidth{0pc} 
\tablecaption{Orbit Model for HD 221673\label{tab::hd221673_parameters}}
\tablehead{
\colhead{Parameter} & \colhead{Value} & \colhead{Uncertainty} 
}
\startdata 
$P_{A-B}$ (days)                 & 179811  & 27745    \\
$T_{A-B}$ (MHJD)                 & 16818   & 3658     \\
$e_{A-B}$                        & 0.322   & 0.047    \\
$a_{A-B}$ (arcseconds)           & 0.568   & 0.065    \\ 
$i_{A-B}$ (deg)              & 21.7    & 8.3      \\
$\omega_{A-B}$ (deg)         & 293     & 15       \\
$\Omega_{A-B}$ (deg)         & 56.2    & 6.0      \\
$P_{\rm Ba-Bb}$ (days)            & 1539    & 51       \\
$T_{\rm Ba-Bb}$ (MHJD)            & 53356   & 32       \\
$e_{\rm Ba-Bb}$                   & 0       & (Fixed)  \\
$a_{COL}$ (${\rm \mu as}$)       & 322     & 29       \\ 
$i_{\rm Ba-Bb}$ (deg)         & 66.6    & 4.0      \\                        
$\omega_{\rm Ba-Bb}$ (deg)    & 0       & (Fixed)  \\
$\Omega_{\rm Ba-Bb, 1}$ (deg) & 128.3   & 4.1       \\
$\chi^2$ and dof                 & 1549.2  & 850       
\enddata
\tablecomments{
Best fit orbital elements in the Campbell basis for HD 221673, with $1\sigma$ 
uncertainties.
}
\end{deluxetable}

\begin{figure}[!ht]
\plotone{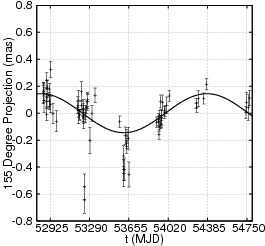}
\caption[Center-Of-Light Orbit of the HD 221673 Subsystem]
{ \label{fig::hd221673_reflex}
Time series of PHASES observations of HD 221673 (72 Peg) for the 1539 day 
subsystem orbital solution, 
measured along an axis at angle $155^\circ$ from 
increasing differential right ascension through increasing differential 
declination; it was along this axis that the PHASES measurements were typically 
most sensitive for this system.  For clarity, only measurements with uncertainties along this 
axis of $100\, {\rm \mu as}$ or less are shown.  
}
\end{figure}

\section{Substellar Companions with Reduced Levels of Confidence}

\subsection{HD 13872}

HD 13872 (21 Ari, HR 657, HIP 10535, WDS 02157$+$2503) is a bright star with 
mid-F dwarf 
spectrum.  In \citeyear{Cou1967a}, it was realized to be a visual binary 
system with separation less than an arcsecond and roughly equal 
luminosities by \citeauthor{Cou1967a}.  Since its first orbit determination, 
there have been questions as to whether its spectral type fit the total system 
mass as measured by the orbit.  \cite{Cou21Ari1982} proposed that an 
additional component must exist in the system to explain the overly large 
total mass.  However, their estimate for the total mass was in error due to 
too small a value of the parallax (of 15 mas).  \cite{Tok1987} pointed out 
that a parallax of 20.1 mas would give a normal sum of masses, a parallax 
later confirmed by {\em Hipparcos} \citep{hipcat}.  However, this is only the 
case if the star's mass ratio is near unity, which had not been determined 
previous to the current investigation, in which a mass ratio 
$M_B/M_A = 1.027 \pm 0.032$ has been measured.

\begin{figure}[!ht]
\plotone{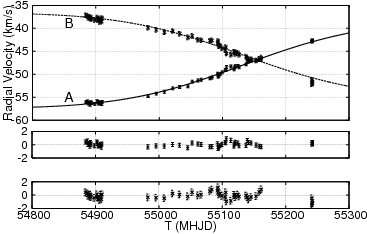}
\caption[Radial Velocity Orbit of 21 Ari]
{ \label{fig::hd13872_rv}
The RV orbit of the 21 Ari binary with measurements from TSU's 
AST.  These measurements enable the binary mass ratio and component masses to 
be measured, but lack the precision necessary to confirm the presence of a 
planetary companion.  The velocities of both components are shown in the top 
graph, where the individual components A and B are labeled.  The middle graph 
shows the measurement residuals for component A, and the bottom graph those 
for component B.
}
\end{figure}

Fifty-one RV measurements of 
HD 13872 were made with Tennessee State 
University's 2 m Automated Spectroscopic Telescope 
\cite[AST; ][]{eatonWilliamsonAST} and echelle 
spectrograph to obtain the stellar mass 
ratio and better constrain the binary orbit.  These measurements are listed in 
Table \ref{tab::HD13872_ast}.  The standard data reduction pipeline for 
determining binary star velocities from AST data was used, as described in 
Paper II.  The RV orbit is plotted in Figure \ref{fig::hd13872_rv}.

\begin{deluxetable}{lllll}
\tablecolumns{5}
\tablewidth{0pc} 
\tablecaption{AST Velocities of 21 Ari \label{tab::HD13872_ast}}
\tablehead{ 
\colhead{Day} 
& \colhead{${\rm RV_A}$} 
& \colhead{$\sigma_{RV, A}$} 
& \colhead{${\rm RV_B}$} 
& \colhead{$\sigma_{RV, B}$}\\
\colhead{(HMJD)} 
& \colhead{(${\rm km \, s^{-1}}$)} 
& \colhead{$({\rm km \, s^{-1}})$}
& \colhead{(${\rm km \, s^{-1}}$)} 
& \colhead{$({\rm km \, s^{-1}})$}}
\startdata
54884.158 & -56.07 & 0.31 & -37.13 & 0.50 \\
54887.098 & -55.91 & 0.31 & -37.31 & 0.50 \\
54888.089 & -56.39 & 0.31 & -37.68 & 0.50 \\
54890.089 & -56.22 & 0.31 & -37.55 & 0.50 \\
54891.094 & -56.63 & 0.31 & -37.83 & 0.50 \\
54892.111 & -56.05 & 0.31 & -37.57 & 0.50 \\
54893.096 & -56.27 & 0.31 & -37.96 & 0.50 \\
54898.123 & -56.45 & 0.31 & -38.16 & 0.50 \\
54902.111 & -55.84 & 0.31 & -37.67 & 0.50 \\
54903.111 & -56.02 & 0.31 & -38.15 & 0.50 \\
54904.125 & -56.40 & 0.31 & -38.06 & 0.50 \\
54906.110 & -56.26 & 0.31 & -38.58 & 0.50 \\
54908.110 & -56.11 & 0.31 & -37.83 & 0.50 \\
54909.117 & -56.48 & 0.31 & -38.54 & 0.50 \\
54910.110 & -56.00 & 0.31 & -37.81 & 0.50 \\
54982.467 & -54.77 & 0.31 & -39.88 & 0.50 \\
54995.464 & -54.21 & 0.31 & -40.33 & 0.50 \\
55008.401 & -53.80 & 0.31 & -40.84 & 0.50 \\
55021.463 & -53.08 & 0.31 & -40.59 & 0.50 \\
55032.425 & -52.77 & 0.31 & -41.16 & 0.50 \\
55045.389 & -52.61 & 0.31 & -41.96 & 0.50 \\
55052.287 & -51.60 & 0.31 & -41.63 & 0.50 \\
55061.360 & -51.35 & 0.31 & -42.48 & 0.50 \\
55066.468 & -51.19 & 0.31 & -42.59 & 0.50 \\
55080.285 & -50.89 & 0.31 & -42.69 & 0.50 \\
55083.447 & -50.71 & 0.31 & -42.61 & 0.50 \\
55092.313 & -49.87 & 0.31 & -43.20 & 0.50 \\
55093.263 & -50.51 & 0.31 & -42.75 & 0.50 \\
55094.434 & -50.26 & 0.31 & -43.87 & 0.50 \\
55096.357 & -49.96 & 0.31 & -43.80 & 0.50 \\
55099.413 & -49.60 & 0.31 & -44.33 & 0.50 \\
55104.390 & -49.10 & 0.31 & -44.73 & 0.50 \\
55105.390 & -49.06 & 0.31 & -44.09 & 0.50 \\
55106.390 & -48.42 & 0.31 & -45.48 & 0.50 \\
55113.390 & -48.31 & 0.31 & -45.75 & 0.50 \\
55118.365 & -48.90 & 0.31 & -45.09 & 0.50 \\
55119.365 & -48.50 & 0.31 & -45.18 & 0.50 \\
55120.365 & -48.25 & 0.31 & -45.56 & 0.50 \\
55124.340 & -48.17 & 0.31 & -45.34 & 0.50 \\
55126.340 & -48.46 & 0.31 & -45.73 & 0.50 \\
55137.315 & -47.84 & 0.31 & -46.13 & 0.50 \\
55139.440 & -47.00 & 0.31 & -46.51 & 0.50 \\
55141.315 & -47.21 & 0.31 & -46.91 & 0.50 \\
55145.290 & -47.08 & 0.31 & -46.75 & 0.50 \\
55153.258 & -47.00 & 0.31 & -47.01 & 0.50 \\
55158.265 & -46.91 & 0.31 & -46.63 & 0.50 \\
55161.240 & -46.84 & 0.31 & -46.38 & 0.50 \\
55241.131 & -42.89 & 0.31 & -51.20 & 0.50 \\
55241.155 & -42.63 & 0.31 & -52.18 & 0.50 \\
55242.131 & -42.60 & 0.31 & -51.61 & 0.50 \\
55242.140 & -42.73 & 0.31 & -52.00 & 0.50 
\enddata
\tablecomments{
Two-component RV measurements 
of 21 Ari (HD 13872) from the AST.
}
\end{deluxetable}

\begin{figure*}[!ht]
\plottwo{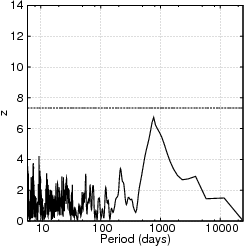}{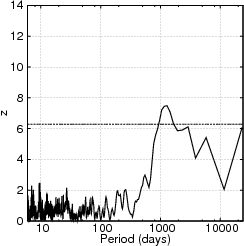}
\caption[Periodograms for Tertiary Companions to HD 13872 (21 Ari)]
{ \label{fig::hd13872_periodograms}
Periodograms of the F statistic comparing models with and without tertiary 
companions to HD 13872 (21 Ari) in astrometric-only models.  The left 
figure is for analysis of only the PHASES data, while the right is for 
combined analysis of PHASES and non-PHASES astrometric measurements.
For HD 13872 the 1\% FAP is at $z=7.33$ for the PHASES-only analysis, and 
$z=6.27$ for the combined analysis, as indicated by horizontal lines.
}
\end{figure*}

The initial PHASES-only search for companions found a most significant peak 
of $z = 6.73$ at a period of $\sim 770$ days with FAP 3.0\%.  This 
relatively low value inspired a 
revised search including both the PHASES and lower precision non-PHASES 
astrometric observations, 
to investigate whether the detection continued to be valid.  The revised 
search finds a most significant peak of $z=7.49$ at a period of $\sim 1284$ 
days with FAP 0.3\%.  The non-zero FAP, especially when only the PHASES 
measurements are considered, prevents identification of this as being counted 
among the strongest of candidates, but is an intriguing possibility, since it 
would correspond to a giant planet.  The periodograms are plotted in Figure 
\ref{fig::hd13872_periodograms}.

\begin{figure*}[!ht]
\plottwo{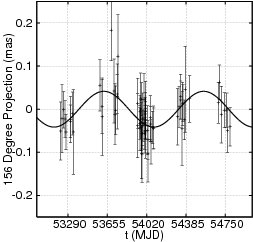}{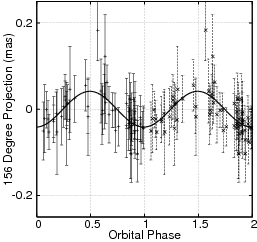}
\caption[Center-Of-Light Orbit of the 21 Ari Subsystem]
{ \label{fig::hd13872_reflex}
Motion of the center of light of the 925 day subsystem in HD 13872 (21 Ari)
measured along an axis at angle $156^\circ$, measured from 
increasing differential right ascension through increasing differential 
declination; it was along this axis that the PHASES measurements were 
typically most sensitive for this binary.  
For clarity, only measurements with uncertainties along this 
axis of $100\, {\rm \mu as}$ or less are shown.  
Left:  the center-of-light motion as a function of 
Modified Julian Date.  Right:  the center-of-light motion phase-wrapped about 
the orbital period and plotted covering two cycles.
}
\end{figure*}

Orbit fitting was refined by allowing the companion orbital period to be 
optimized, non-circular orbits to be considered, and by adding to the data set 
the 51 two-component RV measurements from the TSU AST spectra that span 358 
days, enabling a full three-dimensional orbit to be evaluated, 
including the distance to the system and 
component masses.  There is a small, but insignificant, fit improvement if the 
companion is assumed to be around star A instead of star B.  
The eccentricity of the subsystem orbit is not 
well constrained due to the small signal size and the majority of the 
highest precision astrometry being along only one axis on the sky.  Thus, the 
eccentricity is fixed at zero in the present analysis.  The 2-Keplerian orbit 
model is presented in Table \ref{tab::hd13872_orbit}.  If real, the perturbation 
corresponds to a giant planet of mass $1.40 \pm 0.36 \, \Mjup$.  The reflex motion 
of star A due to the presence of the companion is plotted in 
Figure \ref{fig::hd13872_reflex} with the A-B binary orbit removed.

\begin{deluxetable}{lll}
\tablecolumns{3}
\tablewidth{0pc} 
\tablecaption{Orbit Model for HD 13872\label{tab::hd13872_orbit}}
\tablehead{
\colhead{Parameter} & \colhead{Value} & \colhead{Uncertainty} 
}
\startdata 
$P_{A-B}$ (days)          & 8622.7  & 4.4     \\
$T_{A-B}$ (MHJD)          & 46497.2 & 4.3     \\
$e_{A-B}$                 & 0.68119 & 0.00096 \\
$M_{Aa+Ab}$ ($\Msun$)     & 1.338   & 0.032   \\ 
$M_B$ ($\Msun$)          & 1.374   & 0.027   \\
$i_{A-B}$ (deg)       & 104.437 & 0.025   \\
$\omega_{A-B}$ (deg)  & 263.927 & 0.031   \\
$\Omega_{A-B}$ (deg)  & 55.823  & 0.032   \\
$P_{\rm Aa-Ab}$ (days)    & 925     & 90      \\
$T_{\rm Aa-Ab}$ (MHJD)    & 54092   & 62       \\
$e_{\rm Aa-Ab}$           & 0       & (Fixed)  \\
$M_{Ab}/M_{Aa}$           & 0.00100 & 0.00023  \\
$L_{Ab}/L_{Aa}$           & 0       & (Fixed)  \\
$i_{\rm Aa-Ab}$ (deg) & 71      & 45       \\                        
$\omega_{\rm Aa-Ab}$ (deg) & 0  & (Fixed)  \\
$\Omega_{\rm Aa-Ab, 1}$ (deg) & 211 & 55   \\
$V_{0, {\rm AST}}$ (${\rm km~s^{-1}}$) & -46.892 & 0.053 \\
$d$ (pc)            & 48.90   & 0.33    \\
$\chi^2$ and dof         & 406.1   & 471     
\enddata
\tablecomments{
Best-fit orbital elements in the Campbell basis for HD 13872, with $1\sigma$ 
uncertainties.
}
\end{deluxetable}

The model of HW99 predicts only planets with orbital periods 
less than 210 days will be stable.  This is far smaller than the value of 
$925 \pm 90$ days that best fits the combined measurements.  While the model 
from HW99 is broad in scope for generalized orbits, an analysis specific to 
the configuration of HD 13872 would be beneficial to explore whether there 
are additional islands of stability for companion orbital periods.  However, 
the mutual inclination of the binary and 
planet orbits is not constrained well, limiting 
the utility of system-specific stability analysis.  Clearly this cannot be 
considered a high confidence detection at this time.

Because the data are not of high enough quality to constrain the companion 
eccentricity, the FAP is greater than 0.1\%, and the planet's orbital period 
may be unstable according to the criteria of HW99, the reality of 
this planet is highly uncertain.  HD 13872 
will be an interesting object for continued study, but in this case it might 
not be surprising if future observations do not confirm the presence of a 
giant planet in the system.

\subsection{HD 202444}

HD 202444 ($\tau$ Cyg, 65 Cyg, HR 8130, HIP 104887, WDS 21148$+$3803) is 
classified as an early F subgiant with a G dwarf companion.  Various reports 
have suggested that the primary is a $\delta$ Scuti or $\gamma$ Doradus 
variable, though these have not been confirmed.

\begin{figure*}[!ht]
\plottwo{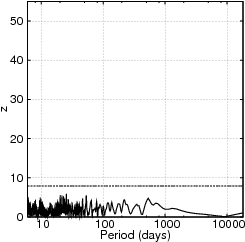}{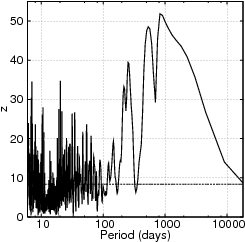}
\caption[Periodograms for Tertiary Companions to HD 202444 ($\tau$ Ceti)]
{ \label{fig::periodograms_202444}
Periodograms of the F statistic comparing models with and without tertiary 
companions to HD 202444 ($\tau$ Ceti) in astrometric-only models.  The left 
figure is for analysis only using the PHASES data, while the right is for 
combined analysis of PHASES and non-PHASES astrometric measurements.
For HD 202444 the 1\% FAP is at $z=8.92$ for the PHASES-only analysis, and 
$z=10.12$ for the combined analysis, as indicated by horizontal lines.
}
\end{figure*}

The search for planetary companions to HD 202444 is more complicated than for 
other stars presented in this paper.  It appears that a companion object may 
exist with an orbital period comparable to the span of PHASES observations 
(only two observations were taken outside of the 1155 day span from MJD 
53234--54389 when most observations of reasonable cadence were made 
whereas the companion orbital period is over 800 days).  Because the 
companion search software reoptimizes both the wide binary orbit model and the 
perturbing model every time a fit is made, the signal could be absorbed into 
that of the wider binary when only the shorter timespan PHASES data were 
analyzed.  Thus, no compelling evidence for a companion was present 
when only PHASES measurements were analyzed---the initial PHASES-only search 
for companions found a most significant peak of $z = 5.93$ at a period of 25.5 
days with FAP 19.1\%.  Analysis of the combined PHASES and 
non-PHASES data sets showed a larger value of $\chi^2$ than one would have 
anticipated based on fits 
to the individual data sets, prompting a second search for tertiary 
companions, this time using all the astrometric measurements.  
When the non-PHASES measurements were added 
to the analysis, the extended coverage of the binary orbit prevented much of 
the ability to adjust the binary orbit to include the perturbations caused by 
possible companions.  The revised 
search finds a very significant peak of $z=51.85$ at a period of 826 days with 
FAP 0.0\%.  The orbit stability criteria of HW99 predict that 
companions with periods shorter than 2200 days are stable in HD 202444, which 
includes all candidate periods identified in the periodograms.  The 
periodograms are plotted in Figure \ref{fig::periodograms_202444}.

The orbit fitting was refined by seeding a full Keplerian fit with the best 
orbital parameters corresponding to the three largest peaks in the full 
data periodogram at periods 826, 534, and 252 days.  The best fit occurred 
for the longest of these periods, for which the eccentricity of the subsystem 
orbit was constrained to be $0.43 \pm 0.17$, so the full Keplerian model 
is accepted.  The final fit has a subsystem period of 810 days, and 
the fit $\chi^2$ is 744.7 with 636 degrees of 
freedom, compared to 994.7 and 643, respectively, for the single Keplerian 
fit.  The best-fit two-Keplerian model parameters are presented in Table 
\ref{tab::hd202444_orbit}.

\begin{figure*}[!ht]
\plottwo{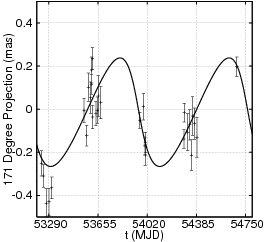}{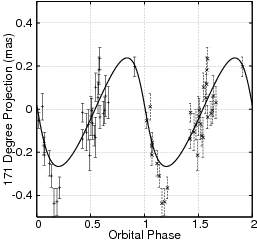}
\caption[Center-Of-Light Orbit of the HD 202444 Subsystem]
{ \label{fig::hd202444_reflex}
Motion of the center of light of the 810 
day subsystem in HD 202444 ($\tau$ Cyg)
measured along an axis at angle $171^\circ$ degrees, measured from 
increasing differential right ascension through increasing differential 
declination; it was along this axis that the PHASES 
measurements were typically 
most sensitive.  For clarity, only measurements with uncertainties along this 
axis of $100\, {\rm \mu as}$ or less are shown.  
Left:  the center-of-light motion as a function of 
Modified Julian Date.  Right:  the center-of-light motion phase-wrapped about 
the orbital period and plotted covering two cycles for continuity.
}
\end{figure*}

Since the companion is not detected when only the PHASES measurements or 
non-PHASES measurements are analyzed individually, there are reasons to 
doubt the authenticity of this proposed companion.  Evidence for the companion 
only appears when the high-precision measurements are 
coupled with the measurements spanning more time.  
The reflex motion of the star in the subsystem is plotted in Figure 
\ref{fig::hd202444_reflex}, in which four measurements with uncertainties 
larger than $100 \, {\rm \mu as}$ projected onto the selected axis 
are not shown; most of 
these points are consistent with this fit, though one suppressed measurement 
at MJD 53285 is a 5.4$\sigma$ outlier.  This measurement was taken with PTI's 
less reliable (and infrequently used) south-west 
baseline, rather than the standard north-south baseline.  
If the companion is a real object, it is at the border between 
the realm of brown dwarfs and giant planets, with a mass of 
$12.3 \pm 2.3 \, \Mjup$, for which a stellar mass of 1.36 $\Msun$ and a 
distance of $20.37 \pm 0.25$ pc have been assumed based on the work of S99.  
In case the candidate 
is not real, the single Keplerian binary-only orbit model was presented in 
Paper II.  Continued high-precision observations of the system 
spanning at least five years would help clarify the situation.

\begin{deluxetable}{lll}
\tablecolumns{3}
\tablewidth{0pc} 
\tablecaption{Orbit model for HD 202444\label{tab::hd202444_orbit}}
\tablehead{
\colhead{Parameter} & \colhead{Value} & \colhead{Uncertainty} 
}
\startdata 
$P_{A-B}$ (days)                 & 18125.4 & 7.7     \\
$T_{A-B}$ (MHJD)                 & 47553   & 17      \\
$e_{A-B}$                        & 0.2392  & 0.0012  \\
$a_{A-B}$ (arcseconds)           & 0.9130  & 0.0013  \\ 
$i_{A-B}$ (deg)              & 134.44  & 0.15    \\
$\omega_{A-B}$ (deg)         & 298.77  & 0.19    \\
$\Omega_{A-B}$ (deg)         & 339.75  & 0.13    \\
$P_{\rm Ba-Bb}$ (days)           & 810     &  18      \\
$T_{\rm Ba-Bb}$ (MHJD)           & 53139   &  48      \\
$e_{\rm Ba-Bb}$                  & 0.43    &  0.17    \\
$a_{COL}$ (${\rm \mu as}$)       & 796    &  149      \\ 
$i_{\rm Ba-Bb}$ (deg)        & 92.6    & 1.9      \\                        
$\omega_{\rm Ba-Bb}$ (deg)   & 90      & 19       \\
$\Omega_{\rm Ba-Bb, 1}$ (deg) & 78.7   &  2.5     \\
$\chi^2$ and dof                & 744.7   & 636      
\enddata
\tablecomments{
Best-fit orbital elements in the Campbell basis for HD 202444, with $1\sigma$ 
uncertainties.
}
\end{deluxetable}

\section{Substellar Companion Candidates with Ambiguous or Uncertain Orbital 
Characteristics}

\subsection{HD 171779}

HD 171779 (HR 6983, HIP 91013, WDS 18339$+$5221) was the first binary for 
which PHASES observations were published as a demonstration of the technique 
\citep{LaneMute2004a}.  In total, 54 differential astrometric measurements 
were made at PTI of this pair of K giant stars.  

\begin{figure*}[!ht]
\plottwo{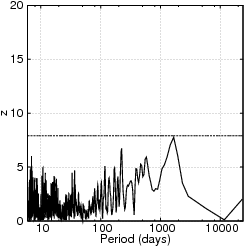}{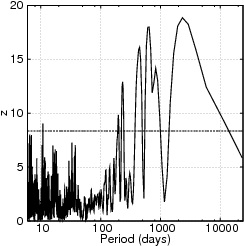}
\caption[Periodograms for Tertiary Companions to HD 171779 (HR 6983)]
{ \label{fig::periodograms_171779}
Periodograms of the F statistic comparing models with and without tertiary 
companions to HD 171779 (HR 6983) in astrometric-only models.  The left 
figure is for analysis of only the PHASES data, while the right is for 
combined analysis of PHASES and non-PHASES astrometric measurements.
For HD 171779 the 1\% FAP is at $z=7.90$ for the PHASES-only analysis, and 
$z=8.36$ for the combined analysis, as indicated by horizontal lines.
}
\end{figure*}

\begin{figure}[!ht]
\plotone{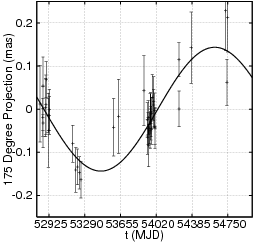}
\caption[Center-Of-Light Orbit of the HD 171779 Subsystem]
{ \label{fig::hd171779_reflex}
Time series of PHASES observations of HD 171779 (HR 6983) for the 2328 day 
subsystem orbital solution, 
measured along an axis at angle $175^\circ$ from 
increasing differential right ascension through increasing differential 
declination; it was along this axis that the PHASES measurements are typically 
most sensitive.  For clarity, only measurements with uncertainties along this 
axis of $100\, {\rm \mu as}$ or less are shown.  
}
\end{figure}

The initial PHASES-only search for companions found a most significant peak 
of $z = 7.76$ at a period of 1663 days with FAP 1.4\%.  This low value 
inspired a 
revised search including both the PHASES and lower-precision non-PHASES 
astrometric observations 
to investigate whether the detection continued to be valid.  The revised 
search finds a most significant peak of $z=18.85$ at a period of 2328 days 
with FAP 0.0\%.  The periodograms resulting from these searches are presented 
in Figure \ref{fig::periodograms_171779}.  
The two approaches find orbital periods within two samples of 
each other at $k = 5$ and $k = 7$, within the $f=3$ oversampling factor of 
the search periods.  Thus, the same signal is detected with both approaches.  
The stability criteria established by HW99 predict orbital 
periods up to 5200 days or more would be stable in this system.  The best-fit 
Keplerian$+$circular subsystem orbital solution is presented in Table 
\ref{tab::hd171779_parameters} and Figure \ref{fig::hd171779_reflex}.

\begin{figure*}[!ht]
\plottwo{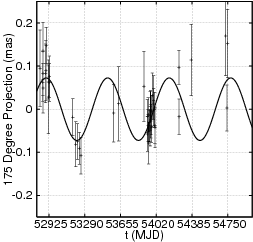}{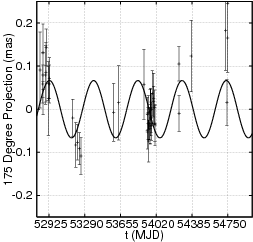}
\caption[Aliased Orbital Periods of the HD 171779 Subsystem]
{ \label{fig::hd171779_timeSeries1}
Time series of PHASES observations of HD 171779 (HR 6983) for two candidate 
orbital periods for companions to HD 171779, showing these solutions 
are likely 
due to cadence rather than an actual object.  The best-fit model for the 
wide A-B binary motion based on a simultaneous fit with the perturbing model 
has been removed in each case, leaving only the motion of the center of light 
of the subsystem.  Measurements are plotted along an axis at angle 
$175^\circ$, measured from increasing differential right ascension through 
increasing differential declination; it was along this axis that the PHASES 
measurements were most sensitive for this binary.  For clarity, only 
measurements with uncertainties along this axis of $100\, {\rm \mu as}$ or 
less are shown.  
Left:  the candidate orbital period is 628 days.
Right:  the candidate orbital period is 451 days.
}
\end{figure*}

However, this is not the only significant peak in the periodogram.  In an 
attempt to uncover which of these might be a true signal, further efforts were 
made to refine the orbital fits with the subsystem orbital period seeded near 
2328 days ($k = 5$), 647 days ($k = 18$), 448 days ($k = 26$), 
831 days ($k = 14$), and 233 days ($k = 50$).  In all of these cases except 
for the 448 day selection, the eccentricity of the subsystem orbit could not 
be constrained by the astrometric measurements; in the case of the 448 day 
period, the best-fit non-zero eccentricity was $0.91 \pm 0.34$, which is only 
poorly defined.  Furthermore, visual inspection of the orbital fit to all 
but the 2328 day perturbation shows the others are almost certainly a result 
of observing cadence rather than a true periodic signal---see Figure 
\ref{fig::hd171779_timeSeries1}.  The 2328 day signal (refined to 2324 days 
after full orbit fitting) is slightly longer than the 1940 day span of 
PHASES observations, and the majority of the measurements might 
be best represented as a 
linear trend, though a circular orbital solution is also possible.  If 
this corresponds to a real companion, the substellar object is either a very 
massive planet or a brown dwarf roughly 
10 times the mass of Jupiter, assuming a distance of 196 pc (based on the 
revised {\em Hipparcos} parallax) and a stellar mass of $1.4 \, \Msun$ 
(derived from the binary orbit).

The multiple peaks in the periodogram, potential for aliasing, inability to 
constrain the eccentricity of the companion, and best-fit orbit period being 
longer than the PHASES observation span prevent this detection from having 
high confidence.  This system warrants further investigation over longer 
timespans to evaluate whether the detected perturbation is real and to 
constrain the orbit.

\begin{deluxetable}{lll}
\tablecolumns{3}
\tablewidth{0pc} 
\tablecaption{Orbit model for HD 171779\label{tab::hd171779_parameters}}
\tablehead{
\colhead{Parameter} & \colhead{Value} & \colhead{Uncertainty} 
}
\startdata 
$P_{A-B}$ (days)          & 75200   & 2464    \\
$T_{A-B}$ (MHJD)          & 21156   & 295     \\
$e_{A-B}$                 & 0.4161  & 0.0083  \\
$a_{A-B}$ (arcseconds)    & 0.2524  & 0.0072  \\ 
$i_{A-B}$ (deg)       &  48.0   & 1.4     \\
$\omega_{A-B}$ (deg)  & 262.4   & 2.9     \\
$\Omega_{A-B}$ (deg)  & 57.5    & 1.3     \\
$P_{\rm Ba-Bb}$ (days)    & 2324    & 250     \\
$T_{\rm Ba-Bb}$ (MHJD)    & 53375   & 21       \\
$e_{\rm Ba-Bb}$           & 0       & (Fixed)  \\
$a_{COL}$ (${\rm \mu as}$) & 160  & 48  \\ 
$i_{\rm Ba-Bb}$ (deg) & 66      & 21       \\                        
$\omega_{\rm Ba-Bb}$ (deg) & 0  & (Fixed)  \\
$\Omega_{\rm Ba-Bb, 1}$ (deg) & 157 & 33   \\
$\chi^2$ and dof         & 335.8   & 352     
\enddata
\tablecomments{
Best-fit orbital elements in the Campbell basis for HD 171779, with $1\sigma$ 
uncertainties.
}
\end{deluxetable}

\subsection{HD 196524}

HD 196524 ($\beta$ Del, 6 Del, HR 7882, HIP 101769, WDS 20375$+$1436) is a 
pair of F5 subgiants.  The system is the 
brightest star in its constellation, despite being given the Bayer designation 
$\beta$.  It and the fainter $\alpha$ Del were given proper names in 
the mid 1800's by Niccol\`o Cacciatore when he compiled the 
Palermo Star Catalogue; $\alpha$ and $\beta$ Del have since been cataloged 
as Sualocin and Rotanev, respectively.  These names are peculiar because they 
are the reverses of Nicolaus and Venator, 
the Latinized versions of Cacciatore's 
own names \citep{allenStarNames}.

\begin{figure*}[!ht]
\plottwo{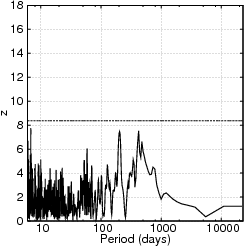}{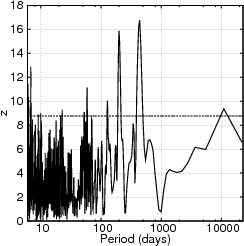}
\caption[Periodograms for Tertiary Companions to HD 196524 (Rotanev/$\beta$ Del)]
{ \label{fig::periodograms_196524}
Periodograms of the F statistic comparing models with and without tertiary 
companions to HD 196524 (Rotanev/$\beta$ Del) in astrometric-only models.  
The left 
figure is for analysis of only the PHASES data, while the right is for 
combined analysis of PHASES and non-PHASES astrometric measurements.
For HD 196524 the 1\% FAP is at $z=8.37$ for the PHASES-only analysis, and 
$z=8.76$ for the combined analysis, as indicated by horizontal lines.
}
\end{figure*}

The initial PHASES-only search for companions found a most significant peak 
of $z = 7.76$ at a period of 6.81 days with FAP 1.9\%.  This value is somewhat 
suspicious since PHASES observations were often scheduled the same nights each 
week.  However, a nearly equal height peak of $z=7.53$ occurs at period 422 
days, and yet another with $z=7.46$ at 203 days.  
The presence of three peaks at very different 
orbital periods, potential for aliasing confusion in the peaks, 
and low FAP values inspired a 
revised search, including both the PHASES and lower-precision non-PHASES 
astrometric observations, to explore the sample of companion orbital periods 
to investigate whether the detection continued to be valid.  The revised 
search finds a most significant peak of $z=16.77$ at a period of 439 days with 
FAP 0.0\%, within one sampling of the secondary peak in the original analysis.  
Furthermore, the peaks at $\sim 200$ and 6.81 days are still present, with 
values of $z=15.88$ and $z=12.88$, respectively.  The periodograms are plotted 
in Figure \ref{fig::periodograms_196524}.

Fits to double Keplerian models were seeded at the three potential companion 
orbital periods to explore which converged on the more satisfactory solution.  
The longest period model converged to a final orbital period of 435 days for 
a circular model with $\chi^2 = 1551.4$ and 1328 degrees of freedom.  The full 
Keplerian version was unable to constrain the eccentricity and the fit failed.  
Analysis of the middle period solution was able to constrain modestly the 
eccentricity to $e = 0.67 \pm 0.24$ with a final $\chi^2 = 1549.1$ and 1326 
degrees of freedom.  Finally, the solution with a period near one week found a 
best-fit period of $6.8116 \pm 0.0016$ days for both circular and eccentric 
($e = 0.27 \pm 0.28$) models, with $\chi^2 = 1570.5$ and $\chi^2 = 1570.2$, 
respectively.  

\begin{figure*}[!ht]
\plottwo{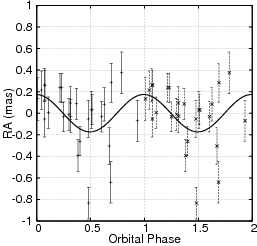}{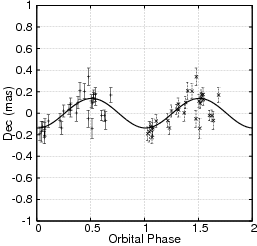}
\caption[The 435 day candidate perturbation to the HD 196524 Subsystem]
{ \label{fig::hd196524_orbit1}
Phase-wrapped PHASES observations of HD 196524 (Rotanev/$\beta$ Del) 
for the longest candidate orbital period for a tertiary companion.  
The best-fit model for the 
wide A-B binary motion based on a simultaneous fit with the perturbing model 
has been removed, leaving only the motion of the center of light 
of the subsystem.  
Left:  motion along the right ascension axis; for clarity, only 
measurements with uncertainties projected on the right ascension axis 
$200\, {\rm \mu as}$ or smaller are shown.  
Right:  motion along the declination axis; for clarity, only 
measurements with uncertainties projected on the declination axis 
$100\, {\rm \mu as}$ or smaller are shown.
}
\end{figure*}

\begin{figure*}[!ht]
\plottwo{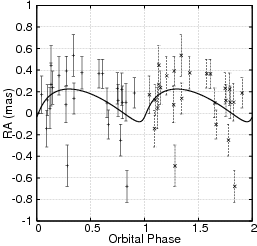}{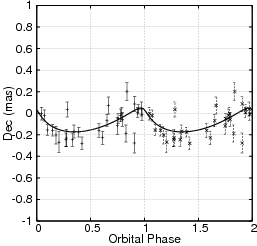}
\caption[The 202 day candidate perturbation to the HD 196524 Subsystem]
{ \label{fig::hd196524_orbit2}
Phase-wrapped PHASES observations of HD 196524 (Rotanev/$\beta$ Del) 
for the 202 day candidate orbital period for a tertiary companion.  
The best-fit model for the 
wide A-B binary motion based on a simultaneous fit with the perturbing model 
has been removed, leaving only the motion of the center of light 
of the subsystem.  
Left:  motion along the right ascension axis; for clarity, only 
measurements with uncertainties projected on the right ascension axis 
$200\, {\rm \mu as}$ or smaller are shown.  
Right:  motion along the declination axis; for clarity, only 
measurements with uncertainties projected on the declination axis 
$100\, {\rm \mu as}$ or smaller are shown.
}
\end{figure*}

Though 96 spectra of this system have been obtained by TSU's AST, the spectral 
features of the two components were blended in all cases and could not be used 
for additional analysis.  However, this does make it less likely that 
the 6.81 day 
signal is evidence of a real companion.  The two longer period solutions are 
presented in Table \ref{tab::hd196524_parameters} and Figures 
\ref{fig::hd196524_orbit1} and \ref{fig::hd196524_orbit2}.  The 
{\em Hipparcos} based parallax of $32.5 \pm 0.7$ mas and average component mass 
of 1.67 $\Msun$ from S99 can be used to convert the stellar reflex 
motion to companion mass.  If real, the 435 
day companion would be a giant planet of $9 \pm 1.6$ 
times the mass of Jupiter whereas the 202 day 
companion would be a brown dwarf of $15 \pm 3.9$ times the mass of Jupiter.

At this point, it is 
not possible to distinguish whether either solution represents an actual 
companion, nor which model is preferred.  Adding to the challenge of evaluating 
this system is the finding that for both solutions the perturbation orbit 
has an orientation on the sky that is more closely aligned with the major axis 
of the typical PHASES uncertainty ellipse.  Where the signal is strongest, the 
astrometric precision is the worst.  HD 196524 is a system that would benefit 
from continued observation by future astrometric efforts, especially those 
capable of truly two-dimensional measurements or that are sensitive to the 
perpendicular axis compared to PHASES.

\begin{deluxetable}{lllll}
\tablecolumns{5}
\tablewidth{0pc} 
\tablecaption{Orbit Models for HD 196524\label{tab::hd196524_parameters}}
\tablehead{
\colhead{Parameter} & \colhead{Value} & \colhead{Uncertainty} & \colhead{Value} & \colhead{Uncertainty} 
}
\startdata 
$P_{A-B}$ (days)          & 9745.6  & 1.4     & 9745.8  & 1.4     \\
$T_{A-B}$ (MHJD)          & 37960.0 & 4.1     & 37961.5 & 4.0     \\
$e_{A-B}$                 & 0.35632 & 0.00070 & 0.35595 & 0.00069 \\
$a_{A-B}$ (arcseconds)    & 0.43676 & 0.00016 & 0.43701 & 0.00016 \\ 
$i_{A-B}$ (deg)       & 61.289  & 0.035   & 61.323  & 0.030   \\
$\omega_{A-B}$ (deg)  & 168.81  & 0.14    & 168.86  & 0.13    \\
$\Omega_{A-B}$ (deg)  & 357.206 & 0.033   & 357.179 & 0.029   \\
$P_{\rm Ba-Bb}$ (days)     & 435.3   & 5.6      & 201.9   & 1.1     \\
$T_{\rm Ba-Bb}$ (MHJD)     & 52941   & 20       & 53013.5 & 7.8     \\
$e_{\rm Ba-Bb}$            &    0    & (Fixed)  & 0.67    & 0.24    \\
$a_{COL}$ (${\rm \mu as}$) & 221    & 40       & 217     & 57      \\ 
$i_{\rm Ba-Bb}$ (deg)  & 87.2    & 4.6      & 84.3    & 3.4     \\
$\omega_{\rm Ba-Bb}$ (deg) & 0  & (Fixed)  & 230     & 23      \\
$\Omega_{\rm Ba-Bb, 1}$ (deg) & 128.3 & 4.0 & 124.2  & 3.9     \\
$\chi^2$ and dof          & 1551.4  & 1328    & 1549.1  & 1326    
\enddata
\tablecomments{
Possible orbits for HD 196524, in the Campbell basis with $1\sigma$ 
uncertainties.
}
\end{deluxetable}

\section{Continued Studies}

Candidate substellar objects discovered by PHASES astrometry include:
\begin{list}{\labelitemi  }{\setlength{\leftmargin}{0.1in}}
\item a planet slightly more massive than Jupiter around one of the 
stars in HD 176051, 
\item one or more brown dwarfs around HD 221673, 
\item a possible Jovian planet orbiting one of the stars in the 
HD 13872 system, though this has low confidence given the non-zero 
FAP of the signal and the prediction that the orbit may not 
be stable over long periods of time, 
\item a possible very massive planet or low mass brown dwarf 
orbiting one of the stars in the HD 202444 binary, though this 
detection has reduced confidence because detection is not possible 
from PHASES measurements alone, but only reveals itself when lower 
precision astrometry covering longer time periods aid in constraining 
the binary orbit, 
\item a possible very massive planet or brown dwarf companion in the HD 
171779 system, though 
at present it is impossible to distinguish which of several possible 
orbital periods are correct, and 
\item a possible massive planet or low-mass brown dwarf companion 
orbiting one of the stars in the HD 196524 system, though which of at 
least two possible orbital periods are correct cannot be determined at 
this time.
\end{list}

Of the 15 binary PHASES targets observed 10 or more times having semimajor 
axis in the 10-50 AU range, Paper III demonstrates the present data set 
cannot rule out planetary mass companions in any stable orbit for 4.  Of the 
remaining 11 systems, there is strong evidence for a Jovian planet companion 
to HD 176051, while HD 13872 may also host a planet, though this detection is 
with lower confidence.  Furthermore, the remaining nine systems in which no 
companions were detected but for which the constraints included some planetary 
mass companions (HD numbers 5286, 76943, 81858, 114378, 137107, 
140436, 202444, 207652, and 214850) 
have a large range of unexplored orbital periods for 
which giant planetary companions cannot be ruled out.  This implies that 
either the PHASES program was incredibly lucky, or giant planets are fairly 
common in close binary systems.  The growing number of such systems being 
detected that are listed in Table \ref{tab:tabclose} suggests that the latter 
explanation is more likely.  

\cite{thebault2004} examined the formation of $\gamma$ Cephei's 
gas giant planet in the core-accretion scenario \citep{mizuno1980}, subject 
to the gravitational perturbations of the binary companion on a moderately 
eccentric ($e = 0.36$) orbit.  Assuming a massive gaseous disk, they 
found that a 10 $M_\oplus$ core could grow in $\sim 10$ Myr, but the core 
always formed at a distance of 1.5 AU, rather than at the observed 2.1 AU.  
Protoplanetary disks are seldom observed to survive for $\sim$ 10 Myr 
around single young stars, much less binary stars, making core accretion 
appear to be an unlikely formation mechanism for gas giants in relatively 
close binary star systems.

The alternative giant planet formation mechanism is disk instability 
\citep{Boss97}. \cite{Nelson2000} modeled disk instabilities in an equal-mass 
binary system with semimajor axis $a = 50$ AU and eccentricity $e = 0.3$, but 
found that the disks became too hot to fragment into gas giant 
protoplanets. On the other hand, \cite{Mayer2005} found that disk 
instabilities could form gas giant planets in binary systems with $e =0.14$ 
and $a = 116$ AU, but with $a = 58$ AU, whether fragmentation 
occurred or not depended on the protoplanetary disk masses and the assumed 
disk cooling rates. \cite{Boss2006} found that disk instabilities 
could lead to giant 
protoplanet formation in binary systems with semimajor axes of 50 or 100 
AU and eccentricities of 0.25 and 0.5. \cite{2007arXiv0705.3182M} tried 
to reconcile these disparate results for disk instability, but could only 
conclude that given the problems with core accretion and the observational 
fact that gas giants exist in binary star systems, disk instability 
remained as a possible formation mechanism for such planetary systems.

The candidate substellar companions discovered by PHASES require continued 
observations by other methods for confirmation.  Because PTI ceased operations 
in 2008, acquiring new PHASES observations will not be possible.  It is 
unlikely any existing northern hemisphere 
long baseline interferometers have the stable astrometric 
baselines required for differential astrometry, though the Navy Prototype 
Optical Interferometer may be a candidate site.  However, it would be better 
if an independent method could be used.  Recent work by 
\cite{helminiak2009} shows 40--1000 ${\rm \mu as}$ precision astrometry using 
adaptive optics (AO) on large telescopes, while 
\cite{lazorenko2007, lazorenko2009} 
show similar precisions without AO on larger fields, 
which in principle might be 
applied to AO images capable of resolving the binaries.  
The PHASES candidate systems 
should be high priority targets for those observing programs; it is 
likely their precisions are sufficient to confirm or reject 
most of the candidate companions.  The SIM-Lite Astrometric Observatory 
\citep{SIM, 2008PASP..120...38U} will also be capable of 
confirming these companions and identifying 
additional systems.

\acknowledgements 
PHASES benefits from the efforts of the PTI collaboration members who have 
each contributed to the development of an extremely reliable observational 
instrument.  Without this outstanding engineering effort to produce a solid 
foundation, advanced phase-referencing techniques would not have been 
possible.  We thank PTI's night assistant Kevin Rykoski for his efforts to 
maintain PTI in excellent condition and operating PTI in phase-referencing 
mode every week.  Thanks are also extended to Ken Johnston and the 
U.~S.~Naval Observatory for their continued support of the USNO Double Star 
Program.  Part of the work described in this paper was performed at 
the Jet Propulsion Laboratory under contract with the National Aeronautics 
and Space Administration.  Interferometer data were obtained at the Palomar
Observatory with the NASA Palomar Testbed Interferometer, supported
by NASA contracts to the Jet Propulsion Laboratory.  This publication makes 
use of data products from the Two Micron All Sky Survey, which is a joint 
project of the University of Massachusetts and the Infrared Processing and 
Analysis Center/California Institute of Technology, funded by the National 
Aeronautics and Space Administration and the National Science Foundation.  
This research has made use of the Simbad database, operated at CDS, 
Strasbourg, France.  M.W.M.~acknowledges support from the Townes Fellowship 
Program, Tennessee State University, and the state of Tennessee through its 
Centers of Excellence program.  Some of the software used for analysis was 
developed as part of the SIM Double Blind Test with support from NASA 
contract NAS7-03001 (JPL 1336910).  
PHASES is funded in part by the California Institute of Technology 
Astronomy Department, and by the National Aeronautics and Space Administration 
under grant no.~NNG05GJ58G issued through the Terrestrial Planet Finder 
Foundation Science Program.  This work was supported in part by the National 
Science Foundation through grants AST 0300096, AST 0507590, and AST 0505366.
M.K.~is supported by the Foundation for Polish Science through a FOCUS 
grant and fellowship, by the Polish Ministry of Science and Higher 
Education through grant N203 3020 35.

{\it Facilities:} \facility{PO:PTI, TSU:AST}

\bibliography{main}
\bibliographystyle{apj}

\end{document}